\documentclass[preprintnumbers,amsmath,amssymbm,prd]{revtex4}
\usepackage{epsfig}
\usepackage{graphicx}
\usepackage{amssymb}

\begin{document}
\title{Bound-state resonances of the Schwarzschild black hole: Analytic treatment}
\author{Shahar Hod}
\affiliation{The Ruppin Academic Center, Emeq Hefer 40250, Israel}
\affiliation{ }
\affiliation{The Jerusalem Multidisciplinary Institute, Jerusalem 91010, Israel}
\date{\today}

\begin{abstract}
\ \ \ Inspired by an earlier idea of Mashhoon, who suggested to relate the discrete quasinormal resonant 
modes of a black hole to the bound-state resonances of the corresponding inverted black-hole potential, 
Völkel [Phys. Rev. Lett. {\bf 134}, 241401 (2025)] has recently computed numerically, for the first time, the bound-state 
energy spectrum of the inverted Schwarzschild potential. 
Motivated by this intriguing work, in the present work we use {\it analytical} techniques 
in order to explore the physical and mathematical properties of the Schwarzschild bound-state resonances. 
In particular, we derive closed-form compact analytical formulas for the 
infinite spectrum $\{E_n\}_{n=0}^{n=\infty}$ of energy 
eigenvalues that characterize the inverted (binding) black-hole potential. 
Interestingly, it is explicitly shown that our analytically derived energy spectrum of the black-hole inverted 
potential agrees remarkably well with the corresponding numerical data that recently appeared in the physics literature. 
\end{abstract}
\bigskip
\maketitle


\section{Introduction}

The dynamics of fundamental fields in curved black-hole spacetimes is characterized by the presence of a relaxation 
phase which is governed by exponentially decaying quasinormal oscillations 
whose discrete complex frequencies $\{\omega_n\}^{n=\infty}_{n=0}$
encode valuable information about the physical properties of the 
central black hole \cite{Nollert1,Press,Cruz,Vish,Davis,QNMs}. 
Quasinormal resonant modes are of fundamental importance in black-hole physics, both from 
observational \cite{Obs1,Obs2} and theoretical perspectives \cite{Hodln3}.

The linearized oscillation modes of spherically symmetric black holes are governed by a Schr\"odinger-like ordinary differential equation of the form \cite{Chan,ThWe,Noteunit,Noterxr}
\begin{equation}\label{Eq1}
\Big[{{d^2}\over{dx^2}}+\omega^2-V(r)\Big]\psi=0\  ,
\end{equation}
where the composed black-hole-field radial potential in (\ref{Eq1}) has the form of 
a scattering (positive) potential barrier. 
The quasinormal oscillation modes, which describe the dissipation of energy into the black hole and 
to spatial infinity, are characterized by the physically motivated boundary conditions of 
ingoing waves at the absorbing horizon of the central black hole and 
outgoing waves at spatial infinity \cite{Detwe}.

Due to the important roles that quasinormal resonant modes play in the physics of black holes, they 
have attracted much attention from both physicists and mathematicians over the last five decades (see 
\cite{Nollert1,Press,Cruz,Vish,Davis,QNMs,Obs1,Obs2,Hodln3} and references therein). 
It should be emphasized that in most situations of physical
interest the complex quasinormal resonant spectra of black holes are not known in closed analytical forms. 
In particular, numerical techniques are usually used in the physics literature in order to compute the 
complex resonant frequencies of composed black-hole-field systems \cite{Nollert1,Press,Cruz,Vish,Davis,QNMs}. 

Mashhoon \cite{Inv1} has intriguingly suggested to relate the quasinormal resonant modes of black holes 
with the bound-state resonances that characterize the corresponding {\it inverted} (negative) black-hole 
potentials (see also \cite{Inv2,Inv3,Inv4,Inv5}). 
Inspired by this interesting idea, Völkel \cite{Vol} has recently used 
numerical computations in order to determine, for the 
first time, the discrete energy levels that characterize the Schr\"odinger-like ordinary differential equation (\ref{Eq1}) with the 
inverted (binding) Schwarzschild black-hole potential $-V(r)$. 

Intriguingly, the recently published important work \cite{Vol} provided compelling numerical evidence that 
the asymptotic (highly-excited) energy levels of the inverted Schwarzschild potential are described extremely well 
by the simple large-$n$ functional behavior 
\begin{equation}\label{Eq2}
E_n={\cal E}(s,l)\cdot e^{-{{2\pi n}\over{\sqrt{l(l+1)-1/4}}}}\ \ \ \ ; \ \ \ \ n\in\mathbb{Z}\  ,
\end{equation}
where ${\cal E}(s,l)$ is a parameter-dependent function 
whose exact functional expression has not been determined in the numerical work \cite{Vol} (here 
$\{s,l\}$ are respectively the spin-weight of the fundamental field \cite{Notes2} and its angular harmonic index).  

The main goal of the present work is to explore, using {\it analytical} techniques, the physical and 
mathematical properties of the infinite set of stationary bound-state resonances that characterize 
the Schr\"odinger-like ordinary differential equation (\ref{Eq1}) with the 
inverted (binding) black-hole potential. 

Below we shall explicitly prove that most of the numerically computed results presented in the 
physically interesting work \cite{Vol} can be derived using purely analytical techniques. 
In particular, we shall present a self-consistent derivation 
of the asymptotic (large-$n$) energy spectrum (\ref{Eq2}) that characterizes 
the inverted Schwarzschild potential. 
Moreover, we shall derive a closed-form compact analytical formula for 
the parameter-dependent energy function ${\cal E}(s,l)$ [see Eq. (\ref{Eq36}) below]. 

\section{Description of the system}

The Schwarzschild black-hole spacetime is characterized by the curved line element \cite{Chan,ThWe,Notesc}
\begin{equation}\label{Eq3}
ds^2=-\Big(1-{{2M}\over{r}}\Big)dt^2+\Big(1-{{2M}\over{r}}\Big)^{-1}dr^2+r^2(d\theta^2+\sin^2\theta
d\phi^2)\  .
\end{equation}
The radius of the black-hole horizon is given by the simple relation 
\begin{equation}\label{Eq4}
r_{\text{H}}=2M\  .
\end{equation}

The dynamics of linearized field modes $\psi(r) e^{-i\omega t}$ in the 
curved black-hole spacetime (\ref{Eq3}) is governed by a Schr\"odinger-like ordinary differential 
equation of the form \cite{Chan,ThWe}
\begin{equation}\label{Eq5}
\Big[{{d^2}\over{dx^2}}+\omega^2-V(r;M,s,l)\Big]\psi=0\  ,
\end{equation}
where the ``tortoise" radial coordinate $x$ is defined by the differential relation \cite{Noterx}
\begin{equation}\label{Eq6}
{{dx}\over{dr}}=\Big(1-{{2M}\over{r}}\Big)^{-1}\  .
\end{equation}
The composed black-hole-field scattering potential in (\ref{Eq5}) is given by the functional 
relation \cite{Notewz,Whee,Zer}
\begin{equation}\label{Eq7}
V(r;M,s,l)=\Big(1-{{2M} \over r}\Big)\Big[{{l(l+1)}\over{r^2}}+{{2M(1-s^2)}\over{r^3}}\Big]\  ,
\end{equation}
where $s$ is the spin-weight of the field ($s=\{0,1,2\}$ for scalar, electromagnetic, and gravitational field 
modes, respectively) and $l\geq s$ is the angular harmonic index of the field mode. 

The infinite spectrum of complex quasinormal resonant frequencies
$\{\omega_n(M,s,l)\}_{n=0}^{n=\infty}$ that characterize the dynamics of linearized fields 
in the curved black-hole spacetime (\ref{Eq3}) is determined by the Schr\"odinger-like ordinary differential 
equation (\ref{Eq5}) with the physically motivated boundary conditions of 
purely ingoing waves at the absorbing horizon of the central black hole and purely outgoing waves at spatial infinity \cite{Detwe}:
\begin{equation}\label{Eq8}
\psi \sim
\begin{cases}
e^{-i\omega x} & \text{ as\ \ \ } r\rightarrow r_{\text{H}}\ \
(x\rightarrow -\infty)\ ; \\ e^{i\omega x} &
\text{ as\ \ \ } r\rightarrow\infty\ \ (x\rightarrow \infty)\  .
\end{cases}
\end{equation}

Following the physically interesting work \cite{Vol} (see also \cite{Inv1,Inv2,Inv3,Inv4,Inv5}) we 
shall study the {\it bound-state} resonances that characterize the 
Schr\"odinger-like ordinary differential equation (\ref{Eq5}) with the {\it inverted} black-hole potential 
\begin{equation}\label{Eq9}
V_{\text{inv}}(r;M,s,l)=-\Big(1-{{2M} \over r}\Big)\Big[{{l(l+1)}\over{r^2}}+{{2M(1-s^2)}\over{r^3}}\Big]\  .
\end{equation}
The effective radial potential (\ref{Eq9}) is negative in the entire
range $x\in(-\infty,\infty)$ and it vanishes asymptotically for
$x\to\pm\infty$. 

Interestingly, we shall explicitly prove below that the Schr\"odinger-like differential equation (\ref{Eq5}) with the 
inverse binding potential (\ref{Eq9}) is amenable to an {\it analytical} treatment. 
In particular, we shall use analytical techniques in order to determine the 
infinite spectrum of bound-state energies $\{E_n\}$ that characterize the non-trivial 
binding potential (\ref{Eq9}).

To this end, it is convenient to define the effective spin index
\begin{equation}\label{Eq10}
p\equiv\sqrt{2-s^2}\
\end{equation}
and the effective harmonic index 
\begin{equation}\label{Eq11}
\ell\equiv-{1\over2}+i\sqrt{l(l+1)-1/4}\
\end{equation}
of the field, in terms of which the Schr\"odinger equation (\ref{Eq5}) with the binding potential (\ref{Eq9}) 
can be written in the form
\begin{equation}\label{Eq12}
{{d^2\psi}\over{dx^2}}+\Big\{\Big[\omega^2-
\Big(1-{{2M}\over{r}}\Big)\Big[{{\ell(\ell+1)}\over{r^2}}+{{2M(1-p^2)}\over{r^3}}\Big]\Big\}\psi=0\  ,
\end{equation}
where the bound-state energies of the system are defined by the relation 
\begin{equation}\label{Eq13}
E=\omega^2<0\
\end{equation}
with $-i\omega\in\mathbb{R}$. 
The bound-state resonances of the system are characterized by asymptotically vanishing (bounded) 
radial eigenfunctions,
\begin{equation}\label{Eq14}
\psi(x\to\pm\infty)\to 0\  .
\end{equation}

\section{The fundamental energy level of the inverted Schwarzschild potential}

In order to calculate the fundamental (lowest) energy level
$E_0(s,l)$ of the inverted (binding) Schwarzschild potential (\ref{Eq9}) we shall closely follow the mathematically 
elegant analysis of Dolan and Ottewill \cite{Dol} who provided a simple analytical technique for the
calculation of the fundamental quasinormal frequencies of composed black-hole-field systems. 
In particular, in this section we shall explicitly demonstrate that the analytical technique of \cite{Dol} 
can also be used to calculate, with a high degree of accuracy, the fundamental {\it energy} level of 
the black-hole inverted potential.

The analytical technique presented in \cite{Dol} is based on an expansion of
the dimensionless resonant frequencies $M\omega_n$ in inverse powers of the dimensionless 
parameter $L\equiv l+1/2$:
\begin{equation}\label{Eq15}
M\omega(s,l;n)=\sum_{k=-1}^{\infty}w_{k}L^{-k}\  ,
\end{equation}
where the first few $\{s,n\}$-dependent expansion coefficients $\{w_k\}$ are given in \cite{Dol}. 

Interestingly, we shall now show that the remarkably simple 
truncated series $M\omega(s,l;n)=\sum_{k=-1}^{1}w_{k}L^{-k}$ yields the fundamental ($n=0$) energy level 
of the inverted binding potential (\ref{Eq9}) with a high degree of accuracy. 
In particular, substituting [see Eq. (\ref{Eq11})] 
\begin{equation}\label{Eq16}
L=\ell+{1\over 2}=i\sqrt{l(l+1)-1/4}
\end{equation}
into Eq. (\ref{Eq15}) with $k\in(-1,0,1)$ and using the relation (\ref{Eq13}) 
one finds the $\{s,l\}$-dependent functional expression \cite{Noteln}
\begin{equation}\label{Eq17}
E_0(s,l)=-{{1}\over{27M^2}}\Bigg[\sqrt{l(l+1)-1/4}-{{1}\over{2}}-
{{{{1}\over{3}}(s^2-1)-{{65}\over{216}}}\over{\sqrt{l(l+1)-1/4}}}\Bigg]^2\
\end{equation}
for the fundamental \cite{Notefun} energy level that characterizes the binding Schwarzschild 
potential (\ref{Eq9}).

In Table \ref{Table1} we present the $l$-dependent 
ratio ${\cal R}_l\equiv {E_0}^{\text{analytical}}/{E_0}^{\text{numerical}}$ 
between the analytically derived fundamental (lowest) energy level of the inverted Schwarzschild potential (\ref{Eq9}) 
[as calculated directly from the analytically derived resonance formula (\ref{Eq17})] and the 
corresponding exact (numerically computed \cite{Vol}) value of the fundamental energy level for gravitational field 
modes with $s=2$. 
Remarkably, the data presented in Table \ref{Table1} for the fundamental bound-state energy levels of the system 
reveals the fact that the agreement between the {\it analytically} derived large-$l$ energy expression (\ref{Eq17}) 
and the corresponding {\it numerically} computed fundamental energy levels of \cite{Vol} is quite 
good (better than $2\%$) in the entire $l\geq2$ regime \cite{Notehl}.

\begin{table}[htbp]
\centering
\begin{tabular}{|c|c|c|c|}
\hline \ \ $l$\ \ & \ $\ 2$\ \ & \ $\ 3$\ \ & \ $\ 4$\ \ \\
\hline \ \ ${\cal R}_l\equiv {{{E_0}^{\text{analytical}}}\over{{E_0}^{\text{numerical}}}}$\ \ &\ \
$0.9811$\ \ \ &\ \ $0.9942$\ \ \ &\ \ $0.9974$\ \ \\
\hline
\end{tabular}
\caption{Fundamental bound-state energies of the inverted (binding) Schwarzschild potential. 
We present, for the case $s=2$ of gravitational field modes, the dimensionless 
ratio ${\cal R}_l\equiv {E_0}^{\text{analytical}}/{E_0}^{\text{numerical}}$ 
between the analytically derived fundamental bound-state energies of the inverted potential [as calculated directly 
from the analytically derived resonance formula (\ref{Eq17})] 
and the corresponding exact values of the energy levels as computed numerically in \cite{Vol}. 
One finds that the agreement between the analytically derived energy formula (\ref{Eq17}) 
and the corresponding numerically computed bound-state energies of \cite{Vol} is remarkably good 
for $l\geq2$.} \label{Table1}
\end{table}

As discussed in \cite{Dol}, the validity of the large-$l$ expansion series 
(\ref{Eq15}) is restricted to the fundamental $n\lesssim l$ modes of the system. 
In the next section we shall develop a different analytical technique 
in order to explore the infinite spectrum of excited energy levels $\{E_n\}$ that characterize the 
Schwarzschild binding potential (\ref{Eq9}).

\section{The infinite spectrum of bound-state energies}

In the present section we shall analyze the {\it infinite} set of excited energy levels $\{E_n\}$ that characterize the 
Schwarzschild inverted (binding) potential (\ref{Eq9}). 
Interestingly, the numerical results presented in \cite{Vol} indicate that the large-$n$ ({\it excited}) energy levels of the system are 
characterized by the strong dimensionless inequalities
\begin{equation}\label{Eq18}
M|\omega_n|\ll1\ \ \ \Longleftrightarrow\ \ \ M^2|E_n|\ll1\ \ \ \ \text{for} \ \ \ \ n\gg1\  .
\end{equation}
As we shall explicitly prove below, a {\it small}-frequency matching analysis [see Eq. (\ref{Eq18})] 
of the ordinary differential equation (\ref{Eq12}) can yield, with a remarkably good degree of accuracy, 
the infinite spectrum $\{E_n\}$ of excited energy levels that characterize the bound-state 
resonances of the inverted black-hole potential (\ref{Eq9}). 

The trick is to analyze the radial Teukolsky equation \cite{TeuPre}
\begin{equation}\label{Eq19}
\Delta^{-p+1}{{d}\over{dr}}\Big(\Delta^{p+1}{{d\psi}\over{dr}}\Big)+\Big[{{\omega^2r^4
-2iM\omega p r^2}}+\Delta[2i\omega pr-\ell(\ell+1)]\Big]\psi=0\ ,
\end{equation}
which, like Eq. (\ref{Eq12}), characterizes the dynamics of linearized spin-weighted fields 
in the curved Schwarzschild black-hole spacetime (\ref{Eq3}) \cite{Noteas}. 
Here 
\begin{equation}\label{Eq20}
\Delta\equiv r^2-2Mr\
\end{equation}
and the effective field parameters $\{p,\ell\}$ 
are given by Eqs. (\ref{Eq10}) and (\ref{Eq11}). 
As explicitly proved by Chandrasekhar \cite{Chanw}, the radial Teukolsky equation
(\ref{Eq19}) for the non-rotating Schwarzschild black hole (also known as the
Bardeen-Press equation \cite{BarPre}) is physically equivalent 
to the Regge-Wheeler equation (\ref{Eq12}).

We are interested in bound-state resonances ($E_n<0$) which are characterized by spatially regular (bounded) 
radial eigenfunctions with the asymptotic properties  
\begin{equation}\label{Eq21}
\psi(x\to -\infty)\sim e^{|\omega|x}\to 0\
\end{equation}
and
\begin{equation}\label{Eq22}
\psi(x\to\infty)\sim e^{-|\omega|x}\to 0\  .
\end{equation}

We shall now prove explicitly that the excited energy spectrum 
$\{E_n\}$ of the Schwarzschild binding potential (\ref{Eq9}) can be 
derived {\it analytically} using a matching procedure in the dimensionless regime (\ref{Eq18}) 
\cite{Notesin,Page,Hodcen}. 
To this end, it is convenient to define the dimensionless variables \cite{Page,Hodcen}
\begin{equation}\label{Eq23}
z\equiv {{r-2M}\over {2M}}\ \ \ \ ; \ \ \ \ k\equiv -2iM\omega\ ,
\end{equation}
in terms of which the radial differential equation (\ref{Eq19}) can be written in the form 
\begin{eqnarray}\label{Eq24}
z^2(z+1)^2{{d^2\psi}\over{dz^2}}+(p+1)z(z+1)(2z+1){{d\psi}\over{dz}}
+\big[-k^2z^4-2pkz^3-\ell(\ell+1)z(z+1)+pk(2z+1)-k^2\big]\psi&=&0\ . \nonumber \\
\end{eqnarray}

The mathematical solution of the differential equation (\ref{Eq24}) in the radial 
region $kz\ll 1$ that satisfies the physically motivated boundary condition (\ref{Eq14})
is given by the functional expression \cite{Page,Hodcen}
\begin{eqnarray}\label{Eq25}
\psi(z)=z^{-p+k}(z+1)^{-p-k} {_2F_1}(-\ell-p,\ell-p+1;1-p+2k;-z)\  , \nonumber \\
\end{eqnarray}
where $_2F_1(a,b;c;z)$ is the hypergeometric function \cite{Abram}.

The mathematical solution of the differential equation (\ref{Eq24}) in the radial region 
$z\gg1$ is given by \cite{Page,Hodcen}
\begin{eqnarray}\label{Eq26}
\psi(z)=A\cdot e^{kz}z^{\ell-p}{_1F_1}(\ell-p+1;2\ell+2;-2kz)+B\cdot e^{kz}z^{-\ell-p-1}{_1F_1}(-\ell-p;-2\ell;-2kz)\ ,
\end{eqnarray}
where $_1F_1(a;c;z)$ is the confluent hypergeometric function \cite{Abram} and 
$\{A,B\}$ are constant coefficients. 

The coefficients $\{A,B\}$ can be determined by matching the two solutions (\ref{Eq25}) and (\ref{Eq26}) 
in the radial overlap region [see Eqs. (\ref{Eq18}) and (\ref{Eq23})] \cite{Noteov}
\begin{equation}\label{Eq27}
1\ll z\ll 1/k\  .
\end{equation}
This matching procedure yields the $\{p,\ell\}$-dependent functional expressions \cite{Page,Hodcen}
\begin{equation}\label{Eq28}
A={{\Gamma(2\ell+1)\Gamma(1-p+2k)}\over{\Gamma(\ell-p+1)\Gamma(\ell+1+2k)}}\
\end{equation}
and
\begin{equation}\label{Eq29}
B={{\Gamma(-2\ell-1)\Gamma(1-p+2k)}\over{\Gamma(-\ell-p)\Gamma(-\ell+2k)}}\ .
\end{equation}

Substituting Eqs. (\ref{Eq28}) and (\ref{Eq29}) into the radial solution (\ref{Eq26}) 
and using the asymptotic ($z\gg1$) properties of the confluent hypergeometric function \cite{Abram}, one
finds the large-$z$ radial functional behavior \cite{Page,Hodcen}
\begin{eqnarray}\label{Eq30}
\psi(z\to\infty)=\psi_1\cdot r^{-2p-1}e^{-kz}+\psi_2\cdot r^{-1}e^{kz}\  ,
\end{eqnarray}
where
\begin{eqnarray}\label{Eq31}
\psi_1={{(2\ell+1)\Gamma^2(2\ell+1)\Gamma(1-p+2k)}\over{\Gamma^2(\ell-p+1)
\Gamma(\ell+1+2k)}}(-2k)^{-\ell-p-1}(2M)^{2p+1} \nonumber \\
-{{(2\ell+1)\Gamma^2(-2\ell-1)\Gamma(1-p+2k)}\over{\Gamma^2(-\ell-p)
\Gamma(-\ell+2k)}}(-2k)^{\ell-s}(2M)^{2p+1}\
\end{eqnarray}
and
\begin{eqnarray}\label{Eq32}
\psi_2={{2(2\ell+1)\Gamma^2(2\ell+1)\Gamma(1-p+2k)}
\over{\Gamma(\ell-p+1)\Gamma(\ell+p+1)\Gamma(\ell+1+2k)}}(2k)^{-\ell+p-1}M \nonumber
\\
-{{2(2\ell+1)\Gamma^2(-2\ell-1)\Gamma(1-p+2k)}\over{\Gamma(-\ell-p)\Gamma(-\ell+p)\Gamma(-\ell+2k)}}
(2k)^{\ell+p}M\  .
\end{eqnarray}

A spatially regular eigenfunction that respects the physically motivated boundary condition
(\ref{Eq14}) is characterized by the asymptotic behavior $\psi(z\to\infty)\to 0$. 
The coefficient $\psi_2$ of the asymptotically diverging exponent in Eq. (\ref{Eq30}) should therefore vanish:
\begin{equation}\label{Eq33}
\psi_2=0\ .
\end{equation}
From Eqs. (\ref{Eq32}) and (\ref{Eq33}) one obtains 
the resonance equation
\begin{equation}\label{Eq34}
(2k)^{2\ell+1}=\Big[{{\Gamma(2\ell+1)}\over{\Gamma(-2\ell-1)}}\Big]^2
{{\Gamma(-\ell-p)\Gamma(-\ell+p)\Gamma(-\ell+2k)}
\over{\Gamma(\ell-p+1)\Gamma(\ell+p+1)\Gamma(\ell+1+2k)}}
\end{equation}
for the bound-state resonances of the Schwarzschild inverted potential (\ref{Eq9}), whose discrete 
solutions in the small-$k$ regime [see Eqs. (\ref{Eq18}) and (\ref{Eq23})] are given by the dimensionless 
functional expression \cite{Notenn}
\begin{equation}\label{Eq35}
k(p,\ell;n)={1\over2}\Big\{\Big[{{\Gamma(2\ell+1)}\over{\Gamma(-2\ell-1)}}\Big]^2
{{\Gamma(-\ell-p)\Gamma(-\ell+p)\Gamma(-\ell)}
\over{\Gamma(\ell-p+1)\Gamma(\ell+p+1)\Gamma(\ell+1)}}\Big\}^{{1}\over{2\ell+1}}\times 
e^{{{-i\pi n}\over{\ell+1/2}}}
\ \ \ \ ; \ \ \ \ n\in\mathbb{Z}\  .
\end{equation}

Taking cognizance of Eqs. (\ref{Eq13}), (\ref{Eq23}), and (\ref{Eq35}) one obtains the resonance expression  
\begin{equation}\label{Eq36}
E(p,\ell;n)=-{1\over16M^2}\Big\{\Big[{{\Gamma(2\ell+1)}\over{\Gamma(-2\ell-1)}}\Big]^2
{{\Gamma(-\ell-p)\Gamma(-\ell+p)\Gamma(-\ell)}
\over{\Gamma(\ell-p+1)\Gamma(\ell+p+1)\Gamma(\ell+1)}}\Big\}^{{1}\over{\ell+1/2}}\times 
e^{-{{2\pi n}\over{\sqrt{l(l+1)-1/4}}}}
\ \ \ \ ; \ \ \ \ n\in\mathbb{Z}\  
\end{equation}
for the infinite spectrum of excited energy levels that characterize the inverted (binding) Schwarzschild potential (\ref{Eq9}) 
in the regime (\ref{Eq18}). 


It is of physical interest to test the accuracy of the analytically derived discrete energy spectrum (\ref{Eq36}), 
which characterizes the excited bound-state resonances of the inverted Schwarzschild potential (\ref{Eq9}), 
against the corresponding exact energy spectrum as computed
numerically in \cite{Vol}. 
In Table \ref{Table2} we present the $n$-dependent (level-dependent) 
ratio ${\cal R}_n\equiv {E_n}^{\text{analytical}}/{E_n}^{\text{numerical}}$ 
between the analytically determined excited energy levels of the inverted Schwarzschild potential (\ref{Eq9}) 
[as calculated directly from the analytically derived large-$n$ resonance formula (\ref{Eq36})] and the 
corresponding exact (numerically computed \cite{Vol}) values of the energy levels for 
gravitational field modes with $l=s=2$. 

The data presented in Table \ref{Table2} for the bound-state resonances of the system 
reveals the fact that the agreement between the {\it analytically} derived energy spectrum (\ref{Eq36}) 
and the corresponding {\it numerically} computed energy spectrum of \cite{Vol} is remarkably good  
in the large-$n$ regime. In fact, the agreement between the analytical and numerical
results is found to be remarkably good (better than $1\%$) already for $n\geq6$.

\begin{table}[htbp]
\centering
\begin{tabular}{|c|c|c|c|c|c|c|c|}
\hline \ \ $n$\ \ & \ $\ 3$\ \ & \ $\ 4$\ \ & \ $\ 5$\ \ & \ $\ 6$\ \ & \ $\ 7$\ \ & \ $\ 8$\ \ & \ $\ 9$\ \ \\
\hline \ \ ${\cal R}_n\equiv {{{E_n}^{\text{analytical}}}\over{{E_n}^{\text{numerical}}}}$\ \ &\ \
$1.1895$\ \ \ &\ \ $1.0495$\ \ \ &\ \ $1.0135$\ \ \ &\ \ $1.0036$\ \ \ &\ \ $1.0009$\ \ \ &\ \ $0.9999$\ \ \ &\ \ $1.0001$\ \ \\
\hline
\end{tabular}
\caption{Bound-state energies of the inverted (binding) Schwarzschild potential. 
We present, for the case $l=s=2$ of gravitational field modes, the dimensionless 
ratio ${\cal R}_n\equiv {E_n}^{\text{analytical}}/{E_n}^{\text{numerical}}$ 
between the analytically determined bound-state energies of the inverted potential [as calculated directly 
from the analytically derived large-$n$ resonance formula (\ref{Eq36})] 
and the corresponding exact values of the energy levels as computed numerically in \cite{Vol}. 
One finds that, in the large-$n$ regime, the agreement between the analytically derived resonance formula (\ref{Eq36}) 
and the corresponding numerically computed resonance spectrum of \cite{Vol} is remarkably good \cite{Notenlw}.} \label{Table2}
\end{table}

\section{Summary}

Quasinormal resonant modes are fundamentally important in black-hole physics, from both 
observational and theoretical perspectives (see 
\cite{Nollert1,Press,Cruz,Vish,Davis,QNMs,Obs1,Obs2,Hodln3} and references therein). 
Mashhoon \cite{Inv1} has raised the intriguing idea to relate the quasinormal resonant modes of black holes 
with the bound-state resonances that characterize the corresponding {\it inverted} black-hole 
potentials (see also \cite{Inv2,Inv3,Inv4,Inv5}). 
Inspired by this suggestion, Völkel \cite{Vol} has recently used numerical computations in order to determine, for the 
first time, the bound-state energy levels of the inverted Regge-Wheeler potential (\ref{Eq9}) that characterizes 
the Schwarzschild black hole. 

Motivated by the interesting numerical results presented in \cite{Vol}, 
in the present work we have studied, using {\it analytical} techniques, the physical and mathematical 
properties of the infinite family of bound-state resonances that characterize 
the Schr\"odinger-like ordinary differential equation (\ref{Eq5}) with the 
inverted (binding) Schwarzschild potential (\ref{Eq9}). 

The main analytical results derived in this work and their physical implications are as follows:

(1) We have demonstrated that the expansion technique (\ref{Eq15}), which originally was
developed in \cite{Dol} for the analysis of black-hole quasinormal frequencies (which are governed by positive 
scattering potentials), 
can also be used with a high degree of accuracy to determine the fundamental energy levels of negative 
binding potentials. 
In particular, we have derived the $\{s,l\}$-dependent functional expression (\ref{Eq17}) for the 
fundamental (lowest) energy level $E_0$ that characterizes the inverted Schwarzschild potential (\ref{Eq9}). 

(2) We have used a small-frequency matching analysis 
in order to solve the Schr\"odinger-like ordinary differential equation (\ref{Eq12}) [with the physically 
motivated boundary conditions (\ref{Eq21}) and (\ref{Eq22})] in the 
dimensionless small-energy $M^2|E_n|\ll1$ regime. 
In particular, we have derived the closed-form compact {\it analytical} formula (\ref{Eq36}) for 
the infinite spectrum $\{E(s,l;n)\}$ of excited energy levels that characterize 
the inverted (binding) Schwarzschild potential (\ref{Eq9}).

(3) It is interesting to note that, defining the arguments 
\begin{equation}\label{Eq37}
\theta\equiv\arg[\Gamma(2\ell+1)]\ \ \ \ ; \ \ \ \ \phi_m\equiv\arg[\Gamma(-\ell+m\cdot p)]\
\end{equation}
and using Eq. 6.1.23 of \cite{Abram}, one can express the analytically derived 
discrete energy spectrum (\ref{Eq36}) in the remarkably compact mathematical form 
\begin{equation}\label{Eq38}
E(s,l;n)=-{1\over16M^2}\cdot\exp\Big[{{4\theta+2\phi_{-1}+2\phi_0+2\phi_1-2\pi n}\over{\sqrt{l(l+1)-1/4}}}\Big]
\ \ \ \ ; \ \ \ \ n\in\mathbb{Z}\  .
\end{equation}

(4) We have explicitly shown (see Tables \ref{Table1} and \ref{Table2}) 
that the {\it analytically} derived formulas (\ref{Eq17}) and (\ref{Eq36}) 
for the bound-state energy spectra of the inverted Schwarzschild potential (\ref{Eq9}) 
agree remarkably well with direct {\it numerical} computations of the corresponding energy levels 
that recently appeared in the physically interesting work \cite{Vol}.  

\bigskip
\noindent
{\bf ACKNOWLEDGMENTS}
\bigskip

This research is supported by the Carmel Science Foundation. I would 
like to thank Yael Oren, Arbel M. Ongo, Ayelet B. Lata, and Alona B. Tea for helpful discussions.


\begin{thebibliography}{99}
\bibitem{Noteunit} We use natural units in which $G=c=\hbar=1$.
\bibitem{Noterxr} The ``tortoise" radial coordinate $x$ in (\ref{Eq1}) is related by a one-to-one correspondence 
to the familiar areal radial coordinate $r$ [see Eq. (\ref{Eq6}) below].
\bibitem{Notesc} Here $(t,r,\theta,\phi)$ are the Schwarzschild spacetime coordinates.
\bibitem{Noterx} Note that the semi-infinite radial regime $r\in[r_{\text{H}},\infty]$ 
is mapped by the differential relation (\ref{Eq6}) into the infinite 
regime $x\in[-\infty,\infty]$.
\bibitem{Notewz} The effective radial potential (\ref{Eq7}) is known as the Regge-Wheeler potential \cite{Whee} 
and it characterizes the dynamics of axial field modes. Polar modes are governed by the Zerilli potential \cite{Zer}. 
It has been established \cite{Chan,ThWe} that these potentials are isospectral.
\bibitem{Noteln} Here we have used the values $\beta\equiv 1-p^2=s^2-1$ and $N\equiv n+1/2=0+1/2=1/2$ ($n=0$ corresponds to the fundamental eigenvalue)
in equations (17)-(19) of \cite{Dol}.
\bibitem{Notefun} Note that the validity of the expansion method
(\ref{Eq15}) is restricted to the regime $n\lesssim l$ \cite{Dol}.
\bibitem{Notehl} Note that higher order terms should be taken in the large-$l$ expansion (\ref{Eq15}) 
in order to achieve an even better accuracy. This is especially true for higher energy levels.
\bibitem{Noteas} Here we have substituted in Eq. (2.9) of \cite{TeuPre} (known as the 
master radial Teukolsky equation) the physical parameters 
$a=0$ for the non-rotating Schwarzschild black hole and 
$\lambda=\ell(\ell+1)-s(s+1)$ for the characteristic angular eigenvalue of the field mode 
in the spherically symmetric spacetime (\ref{Eq3}). 
It is worth noting that our final formula for the infinite spectrum 
of discrete energy levels [see Eq. (\ref{Eq36}) below] is invariant under the transformation $s\to -s$.
\bibitem{Noteov} Note that the radial overlap region $1\ll z\ll1/k$ of the two solutions (\ref{Eq25}) and (\ref{Eq26}) 
exists in the excited energy regime $M^2|E_n|\ll1$ which corresponds to the dimensionless 
small-frequency regime $k\ll1$ [see Eqs. (\ref{Eq13}), (\ref{Eq18}), and (\ref{Eq23})].
\bibitem{Notenn} Here we have used the relation $1=e^{-i2\pi n}$ with $n\in\mathbb{Z}$.
\end{thebibliography}
\end{document}